\documentclass[12pt,preprint]{aastex}       

\usepackage{epsfig,amsmath,natbib}

\usepackage{graphics}

\newcommand{\nh}{$N_{\rm H}$}

\newcommand{\e}[1]{$10^{#1}$}

\newcommand{\cm}[1]{~cm$^{#1}$}
\def\bibi{\bibitem}

\def\beq{\begin{equation}}
\def\enq{\end{equation}}
\def\begar{\begin{eqnarray}}
\def\endar{\end{eqnarray}}
\newcommand{\Msun}{\mbox{$M_{\odot}\;$}}
\def\lsim{\;\raise0.3ex\hbox{$<$\kern-0.75em\raise-1.1ex\hbox{$\sim$}}\;}
\def\gsim{\;\raise0.3ex\hbox{$>$\kern-0.75em\raise-1.1ex\hbox{$\sim$}}\;}

\def\beq{\begin{equation}}
\def\enq{\end{equation}}
\def\begar{\begin{eqnarray}}
\def\endar{\end{eqnarray}}
\def\mathnew{\mathsurround=0pt}
\def\simov#1#2{\lower .5pt\vbox{\baselineskip0pt \lineskip-.5pt
        \ialign{$\mathnew#1\hfil##\hfil$\crcr#2\crcr\sim\crcr}}}

\def\cmc{\rm ~cm^{-3}}

\def\kms{\rm ~km~s^{-1}}
\def\ergs{\rm ~erg~s^{-1}}

\def\etal{{ et al. }}

\def\cmc{\rm ~cm^{-3}}

\def \kms {\rm ~km~s^{-1}}

\def\ergs{\rm ~erg~s^{-1}}

\def\enf{\rm ~erg~cm^{-2}~s^{-1}}

\def \chan {{\it Chandra}}
\def \xmm {{\it XMM-Newton}}

\def\srcXMM{Src 11~}
\def\src{IC~443~}

\def \nh {$N_{\rm H}$ }

\def \hcm {\hbox {\ifmmode $ atom cm$^{-2}\else atom cm$^{-2}$\fi}}
\def \arcmin {\hbox{$^\prime$} }
\def \arcsec {\hbox{$^{\prime\prime}$} }

\def \rchisq {$\chi_{\nu} ^{2}$}
\def\approxgt{\mathrel{\hbox{\rlap{\lower.55ex \hbox {$\sim$}}
        \kern-.3em \raise.4ex \hbox{$>$}}}}
\def\approxlt{\mathrel{\hbox{\rlap{\lower.55ex \hbox {$\sim$}}
        \kern-.3em \raise.4ex \hbox{$<$}}}}
\def\mathnew{\mathsurround=0pt}

\shorttitle{Hard Chandra Source in IC~443}
\shortauthors{Bykov, Bocchino, \& Pavlov}

\begin{document}
\title{
Hard Extended X-ray Source in the IC~443 SNR Resolved by
Chandra:\\
A Fast Ejecta Fragment or a New Pulsar Wind Nebula?}


\author{A.M. Bykov\altaffilmark{1}, F. Bocchino\altaffilmark{2},
and G.G. Pavlov\altaffilmark{3}}
\altaffiltext{1}{A.F. Ioffe Institute of Physics and Technology,
St.\ Petersburg, Russia, 194021; byk@astro.ioffe.ru}
\altaffiltext{2}{INAF-Osservatorio Astronomico di Palermo, Piazza
del Parlamento 1, 90134 Palermo, Italy; bocchino@astropa.unipa.it}
\altaffiltext{3}{Pennsylvania State University, 525 Davey Laboratory,
University Park,
 PA 16802; pavlov@astro.psu.edu}

\begin{abstract}
A \chan\ observation of the isolated hard X-ray source XMMU
J061804.3+222732, located in the region of apparent interaction of
the supernova remnant \src with a molecular cloud, resolved the
complex structure of the source in a few bright clumps embedded in
an extended emission of a $\sim$ 30\arcsec size. The X-ray spectra of
  the clumps and the extended emission are dominated by a hard
power-law component with a photon index of 1.2--1.4.
In addition, we see some indications of an optically thin thermal plasma of a
$\sim$ 0.3 keV temperature. The observed X-ray morphology and
spectra are consistent with those expected for an isolated
supernova ejecta fragment interacting with a dense ambient medium.
A possible alternative interpretation is a pulsar wind nebula
associated with either IC 443 or another SNR, G189.6+3.3.
\end{abstract}



\section{Introduction}

A substantial fraction of core collapsed supernovae are expected
to be spatially correlated with molecular clouds. A well-known
example is IC 443 (G189.1+3.0), an evolved SNR of about
45$\arcmin$ in size, at a distance of 1.5 kpc (e.g., Fesen \&
Kirshner 1980). The presence of a molecular cloud at its boundary
was established from observations of molecular line emission of
OH, CO, and H$_2$, excited by a shock (e.g., Burton et al.\ 1990;
Richter et al.\ 1995). The complex structure of the interaction
region, which consists of multiple dense clumps, is clearly seen
in the Two Micron All Sky Survey (2MASS) images (Rho et al.\
2001). IC~443 is a candidate counterpart of the $\gamma$-ray
source 3EG J0617+2238 detected with EGRET (Esposito et al.\ 1996).
Large-scale soft X-ray mapping, based on the {\sl ROSAT} data
(Asaoka \& Aschenbach 1994), and recent radio observations by
Leahy (2004) suggest that another SNR, G189.6+3.3, is possibly
seen in the IC~443 field.

IC 443 was observed with {\sl HEAO 1\/} (Petre et al.\ 1988), {\sl
Ginga\/} (Wang et al.\ 1992), {\sl ROSAT\/} (Asaoka \& Aschenbach
1994), {\sl ASCA\/} (Keohane et al.\ 1997; Kawasaki et al.\ 2002),
{\sl BeppoSAX\/} (Preite-Martinez et al.\ 1999; Bocchino \& Bykov
2000), and  {\sl RXTE} (Sturner et al.\ 2004).
The {\sl BeppoSAX\/} MECS observations (4--10 keV) identified two
compact sources, 1SAX J0617.1+2221 and 1SAX J0618.0+2227, while
the {\sl BeppoSAX\/} PDS revealed the presence of an unresolved
harder component up to 100 keV. Observations of 1SAX J0617.1+2221
with \chan\
(Olbert et al.\ 2001) and
\xmm\ (Bocchino \&
Bykov 2001) have established the plerionic nature of that source,
while the nature of the other source, 1SAX J0618.0+2227, remained
unknown.

Bocchino \& Bykov (2003; BB03 hereafter) found from
the \xmm\ observations
that the hard emission from
IC 443  is dominated by 12 isolated sources with fluxes
$>5\times 10^{-14}$ $\enf$.
Six of the detected
sources are located in a relatively small, $15^\prime\times
15^\prime$, region. The analysis of the
2MASS
 image
reveals strong 2.2 $\mu m$
emission indicating interaction with a molecular cloud. The source
1SAX J0618.0+2227, the brightest in this region (excluding the
plerion), was resolved with \xmm\ into two sources.
The extended  XMMU J061804.3+222732
and the point-like XMMU J061806.4+222832
(we will call them Src 11 and Src 12, respectively, following BB03)
 have different spectra. The X-ray
spectrum of
Src 11 is hard (photon index $\Gamma \lsim 1.5$).
Src 12 has a steeper featureless spectrum with $\Gamma
\simeq 2.2$. To identify the nature of the sources, we observed
the field with the Advanced CCD Imaging Spectrometer (ACIS) on
board of \chan.


\begin{figure}[tbh]
\begin{center}
      \epsfig{figure=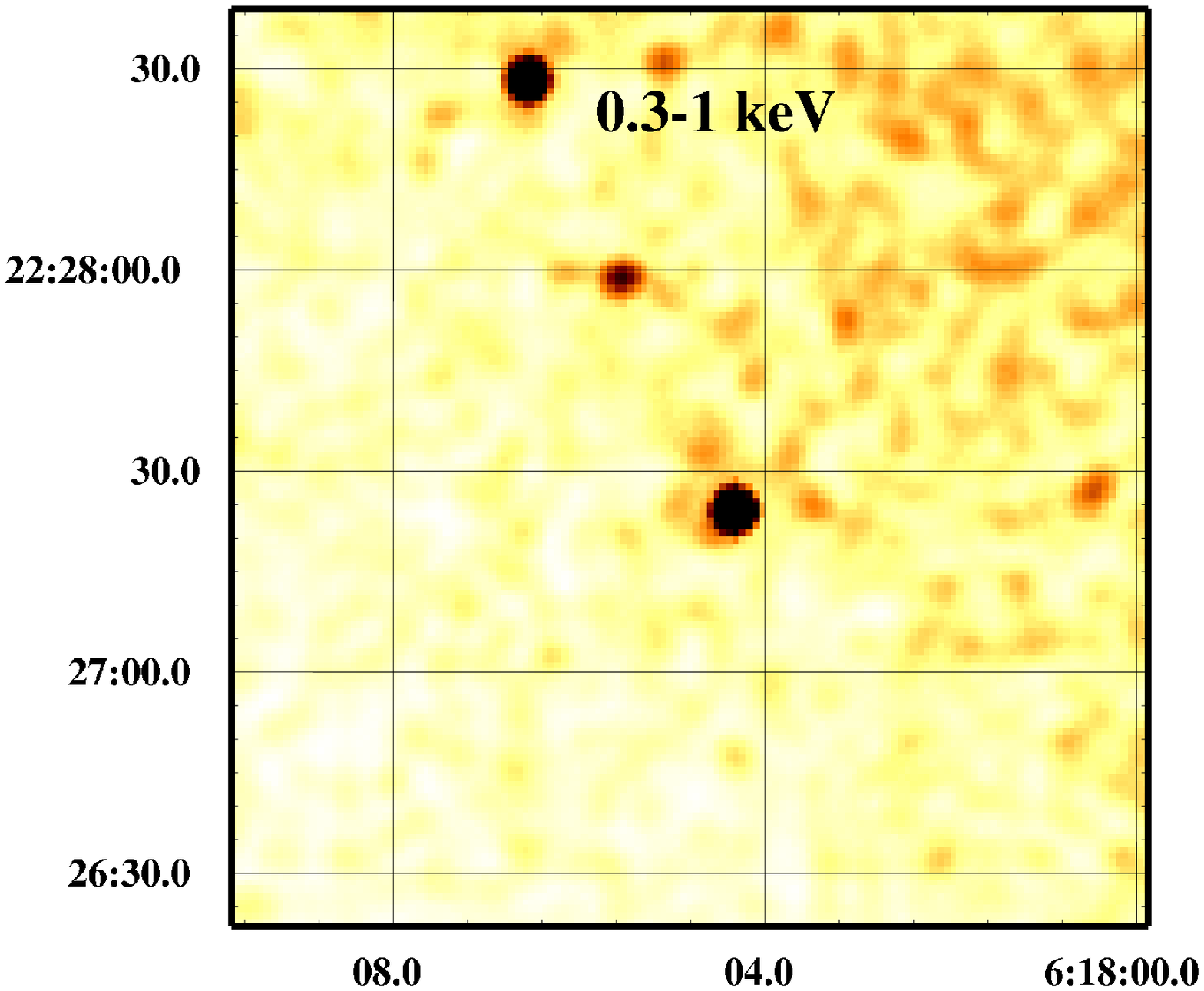,width=7.6cm, bb=40 160 534 623,clip}
      \epsfig{figure=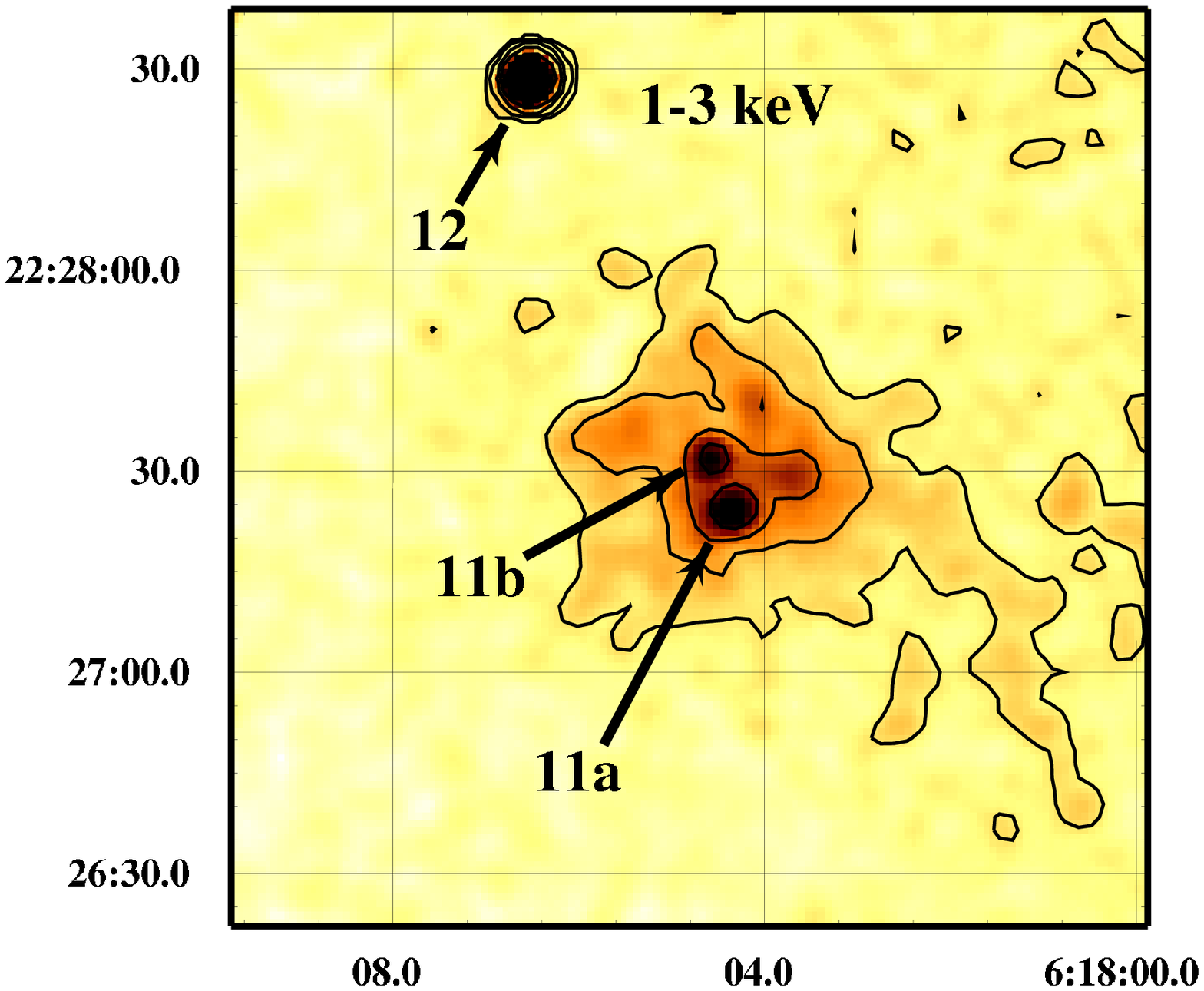,width=7.6cm, bb=40 160 534 623,clip}
      \epsfig{figure=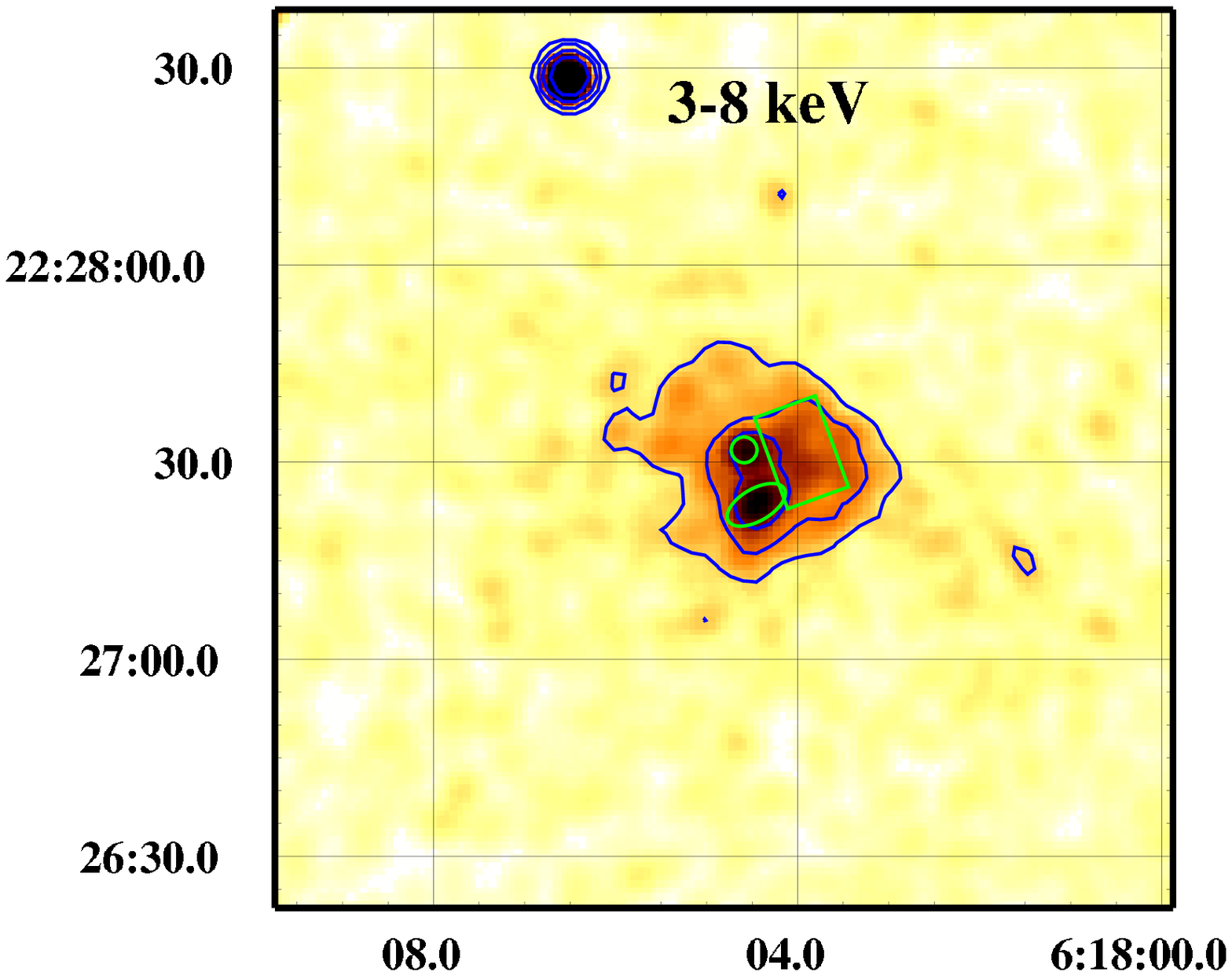,width=7.6cm, bb=40 160 534 623,clip}
\caption{\chan\ ACIS-S3 images of the region of \srcXMM in three
energy bands. The spectral extraction regions used in spectral
analysis are shown in the lower panel.}

\label{fig:image}
\end{center}
\end{figure}
\section{Observations and Data Analysis}
The field of
\srcXMM\ was observed by \chan\ for 59 ksec on
2004 April
12
(ObsID 4675). The source
was imaged at the aim-point
on the ACIS-S3
chip.
 The
event files were processed with CIAO tools (ver.\ 3.1,
CALDB 2.28). To account for the extended structure of the source, the
redistribution matrix files
and weighted auxiliary response
files
were generated using the CIAO tools {\it mkrmf} and
{\it mkwarf}.

\subsection{
Imaging}
Figure~1 shows $2\farcm3\times 2\farcm3$ images,
centered
near the position of Src 11,
in three
 energy bands.
The images, presented in a linear scale, are smoothed with a
Gaussian kernel of a $2''$ width. In the soft energy band, 0.3--1
keV, Src 12 (near the top) and Src 11 (at the center) look like
point sources. In the 1--3 keV image, in addition to the 0.3--1
keV central source (Src 11a hereafter), we see another compact
source, Src 11b, $\approx 8''$ north-north-east of Src 11a,
 as well as patches of diffuse emission surrounding the two sources.
Both Src 11b and the
surrounding diffuse emission
are also
quite prominent in the
3--8 keV image.
Thus, the high-resolution \chan\ images demonstrate that
 the morphology of
\srcXMM\
 is rather complex at energies $\gsim 1$ keV,
showing bright clumps and structured
diffuse emission.

To check whether Src 11a and Src 11b are point-like objects,
we extracted their surface brightness profiles
and compared with a simulated PSF. The positions of the
source centroids were derived using the {\em celldetect}
algorithm provided by CIAO.
 They are
$\alpha = 6^{\rm h}18^{\rm m}04\fs 32$,
$\delta = 22^\circ 27' 24\farcs 1$
and $\alpha = 6^{\rm h}18^{\rm m}04\fs 58$,
$\delta = 22^\circ 27' 31\farcs8$ (J2000.0) for Src 11a and Src
11b, respectively. The errors in the derived positions are
dominated by the uncertainty in the ACIS absolute position
reconstruction, $\simeq 0\farcs 6$ (90\% confidence
level). The PSF was simulated with the {\sl MARX } software (ver. 4.0.8)
using the best-fit spectrum derived by the spectral analysis (see
\S2.2).
 For Src 11a, we found
a significant  deviation
from the model PSF
beyond a 1\farcs5 radial distance (see Fig.\ 2),
indicating an extended structure of the source. We cannot rule out,
however, a possibility
that Src 11a is a soft point source
embedded in a hard extended halo. For
Src 11b, the deviations appear only
beyond $3''$,
so Src 11b resembles a point source on a diffuse background.
There is some indication of a third compact
 source at $\alpha = 6^{\rm h}18^{\rm m}03\fs 39$, $\delta =
22^\circ27'29\farcs5$, which is visible only in the 1--3 keV
 band. A deeper exposure is required to
examine its properties.

\begin{deluxetable}{lccc}
\tablecaption{Fitting parameters for components of Src 11
\label{res}} \tablewidth{0pt} \tablehead{ \colhead{Parameter}  &
\colhead{Src 11a} & \colhead{Src 11b}  & \colhead{R1}
} \startdata

\nh\tablenotemark{a}  & 0.7
                     & 0.7 
                     & 0.7\\ 

$\Gamma$            & 1.3 $\pm$ 0.2
                    & 1.4 $\pm$ 0.2
                    & 1.2 $\pm$ 0.1  \\

$kT$\tablenotemark{b}     & 0.3 $\pm$ 0.1
                          & -- 
                          & -- \\

$F_x$\tablenotemark{c}             & 21.1 & 5.2 & 11.2 \\

$\chi_{\nu}^2$/dof\tablenotemark{d}  & 0.97/17 & 0.91/4 & 0.73/17 \\
\enddata
\tablenotetext{a}{
Hydrogen column density, in \e{22}\cm{-2}, is fixed in the fits.}
\tablenotetext{b}{Temperature for the {\em mekal} model component,
in keV.} \tablenotetext{c}{Unabsorbed
flux in 0.4--7 keV band, in $10^{-14}$ erg cm$^{-2}$ s$^{-1}$.}
\tablenotetext{d}{ The
spectra are grouped
to $\geq$20 source coutns per bin.}
\end{deluxetable}

\subsection{
Spectra}

To investigate the nature of the detected sources, we extracted
spectra from three regions indicated in the
lower panel of
Figure 1:
Src 11a, Src 11b, and the extended rectangular
region R1 in the north-west part of Src 11.
For extraction of background spectra,
we tested several regions
on the ACIS-S3 chip free of point-like
and bright diffuse sources.
The results
were not very sensitive to a particular region choice. The XSPEC
package (ver.\ 11.3) was used to fit
the extracted spectra.

The count statistics for the whole region of
\srcXMM\ is
rather modest
(about 1,000 photons collected).
 The spectral
analysis confirms the
result of the energy-resolved imaging (Fig.\ 1)
that all the structures within the complex Src 11 contain a hard
(power-law) spectral component.
Fits of the brightest Src
11a with single-temperature thermal models
(optically thin plasma or black-body)
are statistically
rather poor (\rchisq\ $>$ 1.4 for 19 dof) or require
unrealistic temperatures above 20
keV.


\begin{figure}[!th]
\begin{center}
      \epsfig{figure=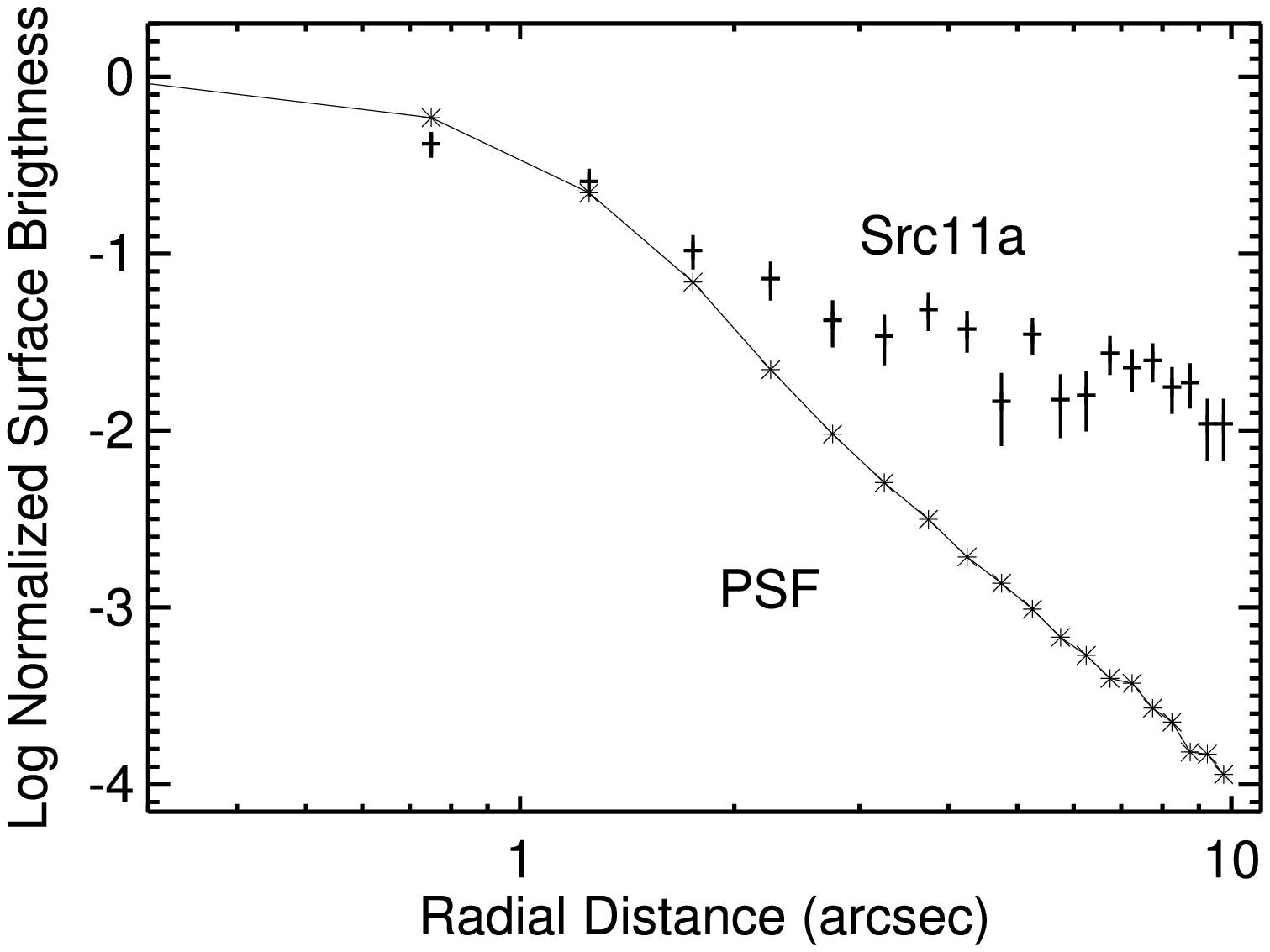,width=8.0cm}
\vspace{-5cm}
\caption{ Radial brightness distribution for Src 11a in the 1--7
keV band and the PSF model simulated with {\sl MARX}.}
\label{fig:psf}
\end{center}
\end{figure}

An absorbed power-law model yields statistically acceptable fits
for each of the spectra of Src 11a, Src 11b, and R1,
with hydrogen column densities $N_{\rm H} = 0.24\pm 0.07$,
$0.66\pm 0.33$, and $0.94\pm 0.33$, respectively
 (in units of $10^{22}$ cm$^{-2}$;
the uncertainties here and in Table 1 correspond to 1 $\sigma$).
 The \nh\ value for Src 11a is well below the absorption column density
$N_{\rm H} \gsim 0.7$
for the south-east region of
IC~443 obtained with {\sl ROSAT} (Asaoka \& Aschenbach 1994).
The \xmm\ data for the spatially unresolved \srcXMM\ (i.e. for all the
three sources combined)  gave
\nh\ = 0.63$^{+0.21}_{-0.16}$ (BB03),
similar to the {\sl ROSAT} fit. We cannot rule out
the possibility that Src 11a
is less absorbed than Src 11b and R1 because it
is a foreground source
(e.g.,
associated with the SNR
G189.6+3.3).
A more plausible interpretation is that
Src 11a, Src 11b, and R1 are
physically
connected, since a chance probability to find two sources with
2--10 keV  fluxes
above $4\times 10^{-14}$ $\enf$ (and
similar photon
indices) within a
$7''$ radius circle in the
anti-centre direction is well below 10$^{-3}$.

If we assume that Src 11a,
Src 11b and R1 are at the same
distance and
fix the column density at \nh\ = 0.7,
the power-law model gives acceptable fits for Src 11b and R1, but
we have to add an additional soft component
for Src 11a (see Fig.\ 3).
Results of such fits, with the additional {\sl mekal}
(optically thin
thermal plasma emission) component for Src 11a, are presented
in Table 1. Such a component is expected if the source is a supersonic
ejecta fragment (see \S3).

A satisfactory fit can also be obtained if we use a blackbody component,
instead of {\em mekal}. Such a blackbody + power-law fit,
with a temperature of $\sim$0.1 keV, radius of $\sim$3 km, and power-law
index of $\sim 1.3$ (at fixed $N_{\rm H} = 0.7$) can be interpreted
as a combination of thermal emission from the surface of a neutron
star plus nonthermal emission from the neutron star magnetosphere
and/or surrounding pulsar wind nebula (PWN).

It is worth noting that
the extrapolation of the hard nonthermal spectrum of Src 11 (see Table 1)
into the EGRET range predicts a photon flux
similar to that of
3EG J0617+2238, whose position is marginally compatible with that of Src 11.
Such a $\gamma$-ray
luminosity can be expected for both the fragment and PWN interpretations.


\begin{figure}[!th]
\begin{center}
      \epsfig{
figure=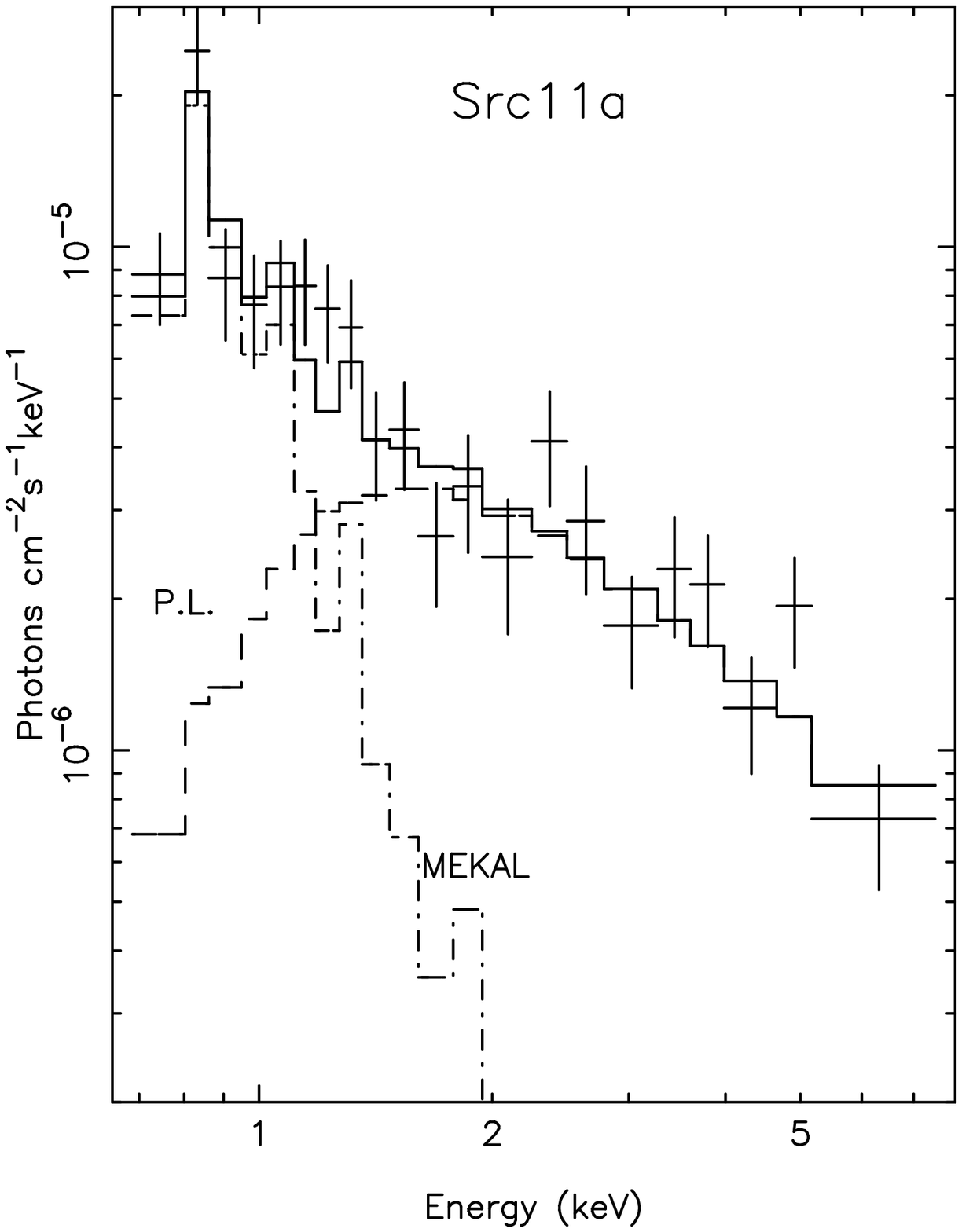,width=8.0cm}
\caption{Spectral fits to Src 11a:
{\em mekal}+power-law model (see Table 1).} \label{fig:src11a_sp}
\vspace{- 0.5cm}
\end{center}
\end{figure}


\section{Discussion}
The analysis presented above reveals the complex structure of Src 11,
projected onto the region where
IC~443 interacts with a molecular cloud.
Below we
discuss
possible identifications of Src 11 with two classes of X-ray objects
relevant to SNRs:
PWNe and isolated ejecta
fragments.

The X-ray source 1SAX J0617.1+2221 in the IC~443 field has been
identified by Olbert et al.\ (2001) and Bocchino \& Bykov (2001)
as a likely PWN. X-ray-radio morphology and the X-ray spectral
steepening toward the outer parts of the nebula support this
identification. Recently, Leahy (2004) speculated that this PWN is
associated with another SNR, G189.6+3.3, rather than IC~443. The
possible presence of two SNRs in the field makes it possible that
Src 11 might also be a low-luminosity ($L_x \sim 10^{32}$  erg
s$^{-1}$ at 1.5 kpc) PWN, possibly powered by a pulsar in Src 11a
that shows a soft thermal-like spectral component. The size ($\sim
0.2$ pc), the hard spectrum, and the presence of compact clumps
are in agreement with this hypothesis. We found no evidence of
spectral steepening in the outskirts of Src 11; however, the count
statistics is scarce, and no substantial synchrotron burning is
expected in a low radiative efficiency PWN. The complex morphology
of Src 11 is rather different from that observed in the 1SAX
J0617.1+2221 PWN. However, if Src 11
 is indeed inside a molecular
cloud, then the PWN appearance can be different from that inside a
hot rarefied SNR gas, and, in fact, many PWNe show rather complex
morphology in high-resolution
images.

Another possible class of compact hard X-ray sources related to
SNRs are isolated ``knots'' associated with fast moving ejecta
fragments. A prototype of
such objects was observed in the Vela SNR (Aschenbach et al.\
1995; Miyata et al.\ 2001). A massive individual fragment moving
supersonically through a molecular cloud would have a luminosity
$L_x \gsim$ 10$^{31} \ergs$ in a 1--10 keV band and would be
observable with \xmm\ and \chan\ from a few kpc distance.

The model of fast supernova fragments predicts two X-ray emission
components (Bykov 2003). The first one is thermal X-ray emission
from a hot shocked ambient gas behind the fragment bow shock, with
the spectrum of an optically thin thermal shocked plasma of an ISM
cloud abundance. A ballistically moving isolated fragment near the
boundary of a SNR of age $t$ would have a velocity
$v \sim R/t$.
For the assumed distance to IC~443 of 1.5 kpc and the age
$t \sim 30$ kyr
suggested by Chevalier (1999), this estimate gives
$v\sim$ 250 $\kms$. The corresponding post-shock gas temperature
is about 0.1 keV,
for a standard estimate $kT = (3/16)\mu m_{\rm H} v^2$. However,
given the uncertainties in the SNR distance and age (e.g., Leahy
2004 suggested that IC~443 is much younger than 30 kyr),
post-shock temperatures of
 about 0.3 keV might be expected as well.
Therefore, our detection of the putative thermal component in the
spectrum of Src 11a is in agreement with this interpretation.

The second emission component is nonthermal. Interaction of the
fast
electrons (accelerated at the fragment bow-shock) with the
fragment body produces both hard continuum and X-ray and IR line
emission, including the K-shell lines of Si and Fe (Bykov 2002).

We apply the model for X-ray  emission from an ejecta fragment of
a 0.2 pc size ($\sim 30\arcsec$ at 1.5 kpc, the current size of
the fragment)
  and a mass of $\gsim 10^{-2} \Msun$,
interacting with a molecular cloud (see Bykov 2002, 2003).
  The knot travels through an
inter-clump medium of a number density $\sim 100 \cmc$ with a
velocity of $\sim 500 \kms$ (corresponding to $kT \sim$ 0.3 keV).
This model predicts a hard continuum with a photon index $\Gamma
\lsim 1.5$ and a normalization consistent with the observed one
(see Table 1). The upper limit on the 1.8 keV Si line flux from
the {\sl Chandra} observation is close to the upper limit $4\times
10^{-6}$ ph cm$^{-2}$ s$^{-1}$ obtained by BB03 with \xmm. To
detect or meaningfully constrain the K-shell  emission lines of Si
and Fe at the level of $\sim 10^{-6}$ ph cm$^{-2}$ s$^{-1}$,
expected for a fragment of Src 11 size and velocity, a longer
\chan\  exposure is required. The Si line would be more suitable
for constraining the nature of Src 11 than the Fe line because
ACIS is more sensitive around 1.8 keV, and the Si line from IC 443
is not absorbed by the ISM, contrary to the case of SNRs in the
Galactic Center region discussed by Bykov (2003).

 Note that to reach its apparent
 position in the molecular cloud,  Src 11 must be massive
enough, $\gsim 10^{-2} \Msun$, to overcome strong drag
deceleration in the dense matter. The probability of another
similarly massive ejecta fragment getting in the ACIS field
of view is small.
 Some of the point-like
X-ray sources seen in the cloud and its vicinity could be smaller
and less massive, $\sim 10^{-3} \Msun$, fragments with luminosities
$L_x \lsim$ 10$^{31} \ergs$ that are
expected to be more numerous.
To firmly identify the sources as the
fragments,
a much deeper observation is required.

The hydrodynamic simulations available (Klein, McKee, \& Colella
1994; Wang \& Chevalier 2002) predict a complex irregular
structure of the fragment body due to the hydrodynamic
instabilities. In this case the X-ray image would show an
irregular patchy structure, instead of the smooth regular
head-tail structure. The patches are due to the emission of dense
pieces of the fragmented knot illuminated by the shock-accelerated
energetic particles. The bright clumps of a few arcsecond size in
Src 11, such as Src 11a and Src 11b, can be explained in that
model.

Thus, the morphology, spectra, and X-ray luminosity of \srcXMM\
are generally consistent with those expected for a ballistically
moving fragment interacting with a molecular cloud. If confirmed
with a deeper exposure, this would provide the first direct
evidence of the potentially important class of hard X-ray sources.

Analysis of IR spectra of Src 11 could help to
distinguish between a massive ejecta fragment and a PWN. The
fragments should be rather numerous in younger SNRs in dense
environments (Bykov 2003). In particular, sources similar to Src
11 can substantially contribute to the unusually rich population
of hard X-ray sources in the Galactic Center region observed with
\chan\ (Muno et al.\ 2003). On the other hand, a reliable
identification of Src 11 with a PWN, e.g. with high resolution
radio observations of Src 11, will be the best test to confirm the
reality of the currently putative SNR G189.6+3.3.

\acknowledgements Support for this work was provided by the NASA
through Chandra Award GO4-5084X and grant NAG5-10865, by RBRF
grants 03-02-17433 and 04-02-16595, and by the International Space
Science Institute (Bern) through the international teams program.




\begin{thebibliography}{}

\bibitem[]{598}
Asaoka, I., \& Aschenbach, B. 1994, A\&A, 284, 573

\bibitem[]{601}
Aschenbach, B., Egger, R. \& Tr\"{u}mper, J. 1995, Nature, 373,
585

\bibi[]{}
Bocchino, F., \& Bykov, A.M. 2000, A\&A, 362, L29

\bibi[]{}
Bocchino, F., \& Bykov, A.M. 2001, A\&A, 376, 248

\bibi[]{}
Bocchino, F., \& Bykov, A.M. 2003, A\&A, 400, 203 (BB03)


\bibi[]{}
Burton, M.G., \etal\
1990, ApJ, 355, 197

\bibi[]{}
Bykov, A.M. 2002, A\&A, 390, 327

\bibi[]{}
Bykov, A.M. 2003, A\&A, 410, L5

\bibi[]{}
Chevalier, R.A. 1999,  ApJ,  511, 798

\bibi[]{}
Fesen, R.A., \& Kirshner, R.P. 1980,  ApJ, 242, 1023


\bibi[]{}
Kawasaki, M.T.  \etal\  2002, ApJ, 572, 897

\bibitem[]{640}
Keohane J.W., et al.
 1997, ApJ, 484, 350


\bibitem[]{644}
Klein, R.I., McKee, C.F.,  \& Colella, P. 1994, ApJ, 420, 213

\bibitem[]{647}
Leahy, D.A. 2004, AJ, 127, 2277


\bibitem[]{651}
Miyata, E., Tsunemi, H., Aschenbach, B. \& Mori, K. 2001, ApJ,
559, L45

\bibitem[]{655}
Muno, M.P. \etal\
2003, ApJ, 589, 225









\bibi[]{}
Olbert, C.M., Clearfield, C.R., Williams, N.E., Keohane, J.W., \&
Frail, D.A. 2001, ApJ, 554, L205


\bibi[]{}
Petre, R. et al. 1988, ApJ, 335, 215

\bibi[]{}
Preite-Martinez A., et al. 1999, AIP Conf. Proc. 510, 73

\bibi[]{}
Richter, M.J., Graham, J.R., \& Wright, G.S. 1995, ApJ,  454, 277

\bibi[]{}
Rho, J. \etal\
 2001,  ApJ, 547, 885.


\bibi[]{}
Sturner, S., Keohane, J., Reimer, O.  2004, Adv. Sp. Res., 33,
429.


\bibitem[]{703}
Wang, C-Y., \& Chevalier, R.A. 2002, ApJ, 574, 155

\end{thebibliography}
\end{document}